\documentstyle[prl,aps,epsf]{revtex}
\begin{document}
\title{A Monopole-Antimonopole Solution 
of the $SU(2)$ Yang-Mills-Higgs Model}
\author{{\large Burkhard Kleihaus}}
\address{\small Department of
Mathematical Physics, National University of Ireland Maynooth}
\author{and}
\address{ } 
\author{{\large Jutta Kunz} }
\address{\small Fachbereich Physik, Universit\"at Oldenburg,
D-26111 Oldenburg, Germany }
\author{ }
\address{ }

\date{\today}
\newcommand{\dd}{\mbox{d}}\newcommand{\tr}{\mbox{tr}}
\newcommand{\ee}{\end{equation}}
\newcommand{\be}{\begin{equation}}
\newcommand{\ii}{\mbox{i}}\newcommand{\e}{\mbox{e}}
\newcommand{\pa}{\partial}\newcommand{\Om}{\Omega}
\newcommand{\bfph}{{\bf \phi}}
\newcommand{\lm}{\lambda}
\newcommand{\st}{\sin\theta}
\newcommand{\ct}{\cos\theta}
\newcommand{\snt}{\sin 2\theta}
\newcommand{\cnt}{\cos 2\theta}

\def\theequation{\arabic{equation}}
\renewcommand{\thefootnote}{\fnsymbol{footnote}}
\newcommand{\re}[1]{(\ref{#1})}
\newcommand{\bfR}{{\sf R\hspace*{-0.9ex}\rule{0.15ex}%
{1.5ex}\hspace*{0.9ex}}}
\newcommand{\N}{{\sf N\hspace*{-1.0ex}\rule{0.15ex}%
{1.3ex}\hspace*{1.0ex}}}
\newcommand{\Q}{{\sf Q\hspace*{-1.1ex}\rule{0.15ex}%
{1.5ex}\hspace*{1.1ex}}}
\newcommand{\C}{{\sf C\hspace*{-0.9ex}\rule{0.15ex}%
{1.3ex}\hspace*{0.9ex}}}
\renewcommand{\thefootnote}{\arabic{footnote}}
\renewcommand{\textfraction}{0.9}

\maketitle
\begin{abstract}

As shown by Taubes, in the Bogomol'nyi-Prasad-Sommerfield limit
the SU(2) Yang-Mills-Higgs model possesses smooth finite energy solutions,
which do not satisfy the first order Bogomol'nyi equations.
We construct numerically such a non-Bogomol'nyi solution,
corresponding to a monopole-antimonopole pair,
and extend the construction to finite Higgs potential.

%\begin{center}
%\Huge DRAFT
%\end{center}
\end{abstract}
 \vfill
 \noindent {Preprint hep-th/9909037} \hfill\break
 \vfill\eject
%\medskip
%\medskip
%\newpage

\section{Introduction}

SU(2) Yang-Mills-Higgs theory, 
with the Higgs field in the adjoint representation,
possesses magnetic monopole and multimonopole solutions.
The solutions with unit magnetic charge are spherically symmetric
\cite{tHooft,Poly,PS,Bog}.
In contrast, multimonopole solutions cannot be spherically symmetric
\cite{ewein} and possess at most axial symmetry
\cite{RR,Ward,Forg,Pras}.
In particular, for magnetic charge greater than two,
solutions with no rotational symmetry exist \cite{monoDS}.

In the limit of vanishing Higgs potential, 
the Bogomol'nyi-Prasad-Sommerfield (BPS) limit,
monopole and multimonopole solutions
satisfy the first order Bogomol'nyi equations \cite{Bogeq}
as well as the second order field equations.
They have minimal energies, saturating precisely the Bogomol'nyi bound.

In the BPS limit,
monopole and axially symmetric multimonopole solutions are known exactly
\cite{PS,Ward,Forg,Pras}.
In contrast, for finite Higgs potential
monopole \cite{tHooft,Bog} and
axially symmetric multimonopole \cite{KKT} solutions
are known only numerically.
But even in the BPS limit,
numerical construction of axial multimonopole solutions \cite{RR}
preceeded their exact construction \cite{Ward,Forg,Pras},
and multimonopole solutions without rotational symmetry
are only known numerically \cite{monoDS}.

As shown by Taubes \cite{Taubes},``there is a smooth, finite action solution
to the SU(2) Yang-Mills Higgs equations in the Bogomol'nyi-Prasad-Sommerfield
limit, which does not satisfy the first-order Bogomol'nyi equations''.
We here construct numerically such a non-Bogomol'nyi BPS solution,
first found in \cite{Rueb}.
This solution possesses axial symmetry
and corresponds to a monopole-antimonopole pair.
We extend the construction to finite Higgs potential.

We review the SU(2) Yang-Mills-Higgs model in section II
and the axially symmetric ansatz for the monopole-antimonopole solution
in section III.
We analyze the magnetic charge of the solution in section IV
and present numerical results in section V.
In section VI we present the conclusions.

\boldmath
\section{$SU(2)$ Yang-Mills-Higgs Model}
\unboldmath

We consider the $SU(2)$ Yang-Mills-Higgs Lagrangian
\begin{eqnarray}
-{\cal L} & = & 
\int\left\{
 \frac{1}{2g^2} Tr\left( F_{\mu\nu} F^{\mu\nu}\right)
+\frac{1}{4} Tr\left(D_\mu \Phi D^\mu \Phi\right)
+\frac{\lambda}{2} Tr\left((\Phi^2-\eta^2)^2\right)
\right\} d^3r
\label{lag1}
\  \end{eqnarray}
with field strength tensor of the $su(2)$ gauge potential 
$A_\mu = \frac{1}{2} \tau_a A_\mu^a$,
\begin{eqnarray}
F_{\mu\nu} & = & \partial_\mu A_\nu - \partial_\nu A_\mu 
                  + i \left[A_\mu, A_\nu \right] \ ,
\label{Fdef}
\end{eqnarray}
and covariant derivative
of the Higgs field $\Phi = \tau_a \phi^a$ in the adjoint representation
\begin{eqnarray}
D_\mu \Phi & = & \partial_\mu \Phi + i \left[ A_\mu, \Phi \right] 
\label{DPdef} \ , \end{eqnarray}
and $g$ denotes the gauge coupling constant,
$\lambda$ the strength of the Higgs potential and
$\eta$ the vacuum expectation value of the Higgs field.

The Lagrangian (\ref{lag1}) is invariant under local $SU(2)$ 
gauge transformations ${\bf g}$, 
\begin{eqnarray}
A_\mu &\longrightarrow & {\bf g} A_\mu {\bf g}^{-1} 
+ i \partial_\mu {\bf g} {\bf g}^{-1} \ ,
\nonumber\\
\Phi  &\longrightarrow & {\bf g} \Phi {\bf g}^{-1} \ .
\nonumber\\
\end{eqnarray}

Static finite energy configurations can be characterized 
by an integer topological charge $Q$
\begin{equation}
Q = \frac{1}{4\pi\eta} \int{Tr\left\{F_{ij} D_k\Phi\right\}
 \varepsilon^{ijk}} d^3r 
\label{topN}
\ , \label{Q} \end{equation}
corresponding to the magnetic charge $m=Q/g$.
In the BPS limit
the energy $E$ of configurations with topological charge $Q$
is bounded from below
\begin{equation}
E \ge \frac{4 \pi \eta Q}{g}
\ . \label{E} \end{equation}
Monopole and multimonopole solutions
satisfying the Bogomol'nyi condition
\begin{equation}
F_{ij}^a = \varepsilon_{ijk} D_k \phi^a
\   \label{Bogc} \end{equation}
precisely saturate the lower bound (\ref{E}).

Here we construct a solution, which corresponds to
a monopole-antimonopole configuration and therefore
carries $Q=0$. It has finite energy $E>0$,
and thus, in the BPS limit, corresponds to
a non-Bogomol'nyi solution of the $SU(2)$
Yang-Mills-Higgs field equations.

\boldmath
\section{Static axially symmetric $Q=0$ Ansatz}
\unboldmath

We choose the static, axially symmetric, purely magnetic Ansatz 
employed in \cite{Rueb} for the monopole-antimonopole solution
and in \cite{Klink,KuBri} for the sphaleron-antisphaleron
solution of the Weinberg-Salam model.
Here the gauge field is parametrized by
\begin{equation}
A_0 = 0\ , 
\ \ A_r =\frac{H_1}{2gr}\tau_\varphi\ , 
\ \ A_\theta = \frac{\left(1-H_2\right)}{g} \tau_\varphi, 
\ \ A_\varphi = -\frac{\sin\theta}{g}\left(H_3 \tau^{(2)}_r+
\left(1-H_4\right)\tau^{(2)}_\theta\right)\  \ ,
\label{A_an}
\end{equation}
and the Higgs field by
\begin{equation}
\Phi =\eta
\left( \Phi_1 \tau^{(2)}_r \ + \Phi_2 \tau^{(2)}_\theta\right) \ .
\label{Phi_an}
\end{equation}
The $su(2)$ matrices 
$\tau^{(2)}_r$, $\tau^{(2)}_\theta$ and $\tau_\varphi$
are defined in terms of the Pauli matrices $\tau_1,\tau_2,\tau_3$ by
\begin{eqnarray}
\tau^{(2)}_r & = & \sin 2\theta(\cos\varphi \tau_1 + \sin\varphi \tau_2) 
             + \cos 2\theta \tau_3 \ , 
\nonumber\\             
\tau^{(2)}_\theta & = & \cos 2\theta(\cos\varphi \tau_1 + \sin\varphi \tau_2)
             - \sin 2\theta \tau_3 \ , 
\nonumber\\             
\tau_\varphi & = & -\sin\varphi \tau_1 + \cos\varphi \tau_2 \ ,
\end{eqnarray}
and for later convenience we define 
\begin{equation}
\tau_\rho  =  \cos\varphi \tau_1 + \sin\varphi \tau_2 \ .
\nonumber
\end{equation}

We change to dimensionless coordinates and Higgs field by rescaling 
$r \rightarrow r/(g \eta)$ and $\Phi\rightarrow \eta \Phi$, respectively.
Then this Ansatz leads to the field strength tensor 
\begin{eqnarray}
F_{r\theta} & = & 
 -\frac{1}{2 r}\left( \partial_\theta H_1 + 2 r \partial_r H_2\right)
                   \tau_\varphi \ ,
\nonumber\\
F_{r\varphi}  & = &  \frac{1}{2 r}\left\{
            \left( \snt H_1 - 2\st  H_1 (1-H_4) -2 \st r \partial_r H_3\right)
            \tau^{(2)}_r 
            \right.
\nonumber\\ & & 
            \left.
           + \left( \cnt H_1 + 2\st  H_1 H_3     +2 \st r \partial_r H_4\right)
           \tau^{(2)}_\theta  
            \right\} \ ,
\nonumber\\
F_{\theta\varphi}  & = & -\frac{1}{2}\left\{
         \left(2 \snt (H_2-1) +2 \ct H_3 -2\st H_2 (1-H_4) 
                                    +2 \st \partial_\theta H_3\right)
            \tau^{(2)}_r
            \right.
\nonumber\\ & & 
            \left.
        +\left(2 \cnt (H_2-1) +2 \ct (1-H_4) +2\st H_2 H_3
                                    -2 \st \partial_\theta H_4\right)
            \tau^{(2)}_\theta
            \right\} \ ,
\label{Fij}
\end{eqnarray}
and the covariant derivative of the Higgs field 
\begin{eqnarray}
D_r\Phi  & = & \frac{1}{r}\left\{
                   \left( r \partial_r \Phi_1+ H_1 \Phi_2 \right)  
                            \tau^{(2)}_r
                  +\left( r \partial_r \Phi_2- H_1 \Phi_1 \right)  
                            \tau^{(2)}_\theta  
              \right\} \ ,
\nonumber\\
D_\theta\Phi  & = & \left(\partial_\theta \Phi_1 - 2 H_2 \Phi_2\right) 
                            \tau^{(2)}_r
                   +\left(\partial_\theta \Phi_2 + 2 H_2 \Phi_1\right)                             
                            \tau^{(2)}_\theta \ ,
\nonumber\\
D_\varphi\Phi  & = & \left\{\left(\snt -2 \st (1-H_4)\right)\Phi_1 
                   +\left(\cnt +2 \st H_3\right) \Phi_2\right\}
                            \tau_\varphi \ .
\label{coD}
\end{eqnarray}
The dimensionless energy density then becomes 
\begin{eqnarray}
\varepsilon & = & 
{\rm Tr} \left\{ 
\frac{1}{r^2}F_{r\theta}^2 
+\frac{1}{r^2\sin^2\theta}F_{r\varphi}^2 
+\frac{1}{r^4\sin^2\theta}F_{\theta\varphi}^2 
\right\}
\nonumber\\ 
& & 
+\frac{1}{4}
{\rm Tr} \left\{
(D_r\Phi)^2
+ \frac{1}{r^2}(D_\theta\Phi)^2
+ \frac{1}{\sin^2\theta^2}(D_\varphi\Phi)^2
\right\}
+\lambda
(\left(|\Phi|^2-1\right)^2 \ ,
\end{eqnarray}
where $|\Phi|=\sqrt{\Phi_1^2+\Phi_2^2}$ denotes the modulus of the Higgs field.

For finite energy configurations the modulus of the Higgs field has 
to be one at infinity, whereas the covariant derivatives of the 
Higgs field have to vanish at infinity. These conditions lead to
\cite{Rueb}
\begin{equation}
 r\longrightarrow \infty \ :  
\Phi_1 \longrightarrow  1 \ , \ \ \Phi_2  \longrightarrow  0 \ ,
\label{asymp1}
\end{equation}
\begin{equation}
 r\longrightarrow \infty \ :  
H_1  \longrightarrow  0 \ , \ \ 
H_2  \longrightarrow  0 \ , \ \ 
H_3  \longrightarrow  \st \ , \ \ 
1-H_4  \longrightarrow  \ct \ .
\label{asymp2}
\end{equation}
Substituting the asymptotic expressions for the gauge field functions
into the Ansatz (\ref{A_an}) 
shows, that the gauge potential approaches a pure gauge at infinity 
\begin{eqnarray}
A_r & \longrightarrow &\ \ \ \ \ \ \ \ \ \ \ \ \ \ \ 0 \ ,
\nonumber\\
A_\theta &\longrightarrow & 
\ \ \ \ \ \ \ \ \ \ \ \ \ \ \ \tau_\varphi\ \ \ \ \ \ \ \ \ \ \ \ \ \ \ \ \  
               = -i \partial_\theta {\bf g} {\bf g}^\dagger \ ,
\nonumber\\
A_\varphi &\longrightarrow & -\st (\ct \tau_\rho + \st \tau_\varphi ) \ \ \
               = -i \partial_\varphi {\bf g}{\bf g}^\dagger \ ,
\end{eqnarray}
where ${\bf g}=\exp\{i \theta \tau_\varphi\}$
rotates the Higgs field at infinity to a constant,
${\displaystyle \left.\left({\bf g} 
  \Phi {\bf g}^\dagger\right) \right|_\infty= \tau_3 }$
\cite{note1}.
Reexpressing the topological charge (\ref{Q}) 
as a surface integral, we find $Q=0$ for configurations 
obeying (\ref{asymp1}), (\ref{asymp2}),
provided, the configurations are sufficiently regular.

Inserting the Ansatz (\ref{A_an}), (\ref{Phi_an}) 
into the general variational equations 
leads to a system of six coupled non-linear partial differential equations
for the four gauge field functions $H_i$ and 
the two Higgs field functions $\Phi_i$.
The same system of partial differential equations
is obtained by inserting the Ansatz directly into the Lagrangian 
and calculating the variation with respect to the functions $H_i$ and 
$\Phi_i$, showing that
the Ansatz (\ref{A_an}), (\ref{Phi_an}) is self-consistent.

The Ansatz is not a priori well defined on the $z$-axis and at the origin.
However, for solutions of the field equations, 
we have performed an expansion of the gauge and
Higgs field functions near the $z$-axis and near the origin
\cite{Regu}. Inserting these
expansions into the Ansatz we find that the gauge potential and the 
Higgs field are well defined and  
(at least) twice differentiable on the $z$-axis and at the origin. 

At the origin we find
\begin{eqnarray}
A_x & = & -\frac{g_3}{2}x y \tau_1
     + \left[\frac{z}{2}(g_1+2 g_2) +\frac{g_3}{4}(2x^2-z^2)\right] \tau_2
     -g_4 y z \tau_3 \ ,
\nonumber\\           
A_y & = &  -\left[\frac{z}{2}(g_1+2 g_2) +\frac{g_3}{4}(2y^2-z^2)\right] \tau_1
      +\frac{g_3}{2}x y \tau_2
      +g_4 x z \tau_3 \ ,
\nonumber\\           
A_z &  =& (g_2+g_3 z)(y \tau_1 -x \tau_2)  \ ,     
\nonumber\\           
\Phi & = &\left[\frac{g_3 g_5}{10}(3 \rho^2-7 z^2)
                       -g_6(\rho^2+4 z^2)\right] (x\tau_1+y\tau_2)
\nonumber\\           
 & &
       +\left[\phi_0 -4 \phi_0 \lambda z^2 (1-\phi_0^2) 
         +z (\frac{\phi_0 g_5}{5} -g_6) (3\rho^2-2 z^2)
             +g_7(\rho^2 - 2 z^2)\right]\tau_3  \ ,
\end{eqnarray}
where $\phi_0, g_i$ are constants.
Therefore the Ansatz allows for 
a non-vanishing Higgs field, $\Phi(r=0)=\phi_0 \tau_3$,
at the origin.

Near the $z$-axis the gauge field functions behave like
\begin{eqnarray}
H_1 & = & h_{11}(r)\st+\dots \ , \ \ 
H_2 = f(r) + h_{22}(r) \sin^2\theta +\dots\ , 
\nonumber\\
H_3 & = & h_{31}(r) \st +\dots\ , \ \ 
H_4 = f(r) + h_{42}(r) \sin^2\theta +\dots \ , 
\end{eqnarray}
while the Higgs field functions behave like
\begin{equation}
\Phi_1 = \phi(r) + \phi_{12}(r) \sin^2\theta +\dots \ , \ \ 
\Phi_2 = \phi_{21}(r) \st +\dots \ , 
\end{equation}
where $\dots$ indicate higher order terms in $\sin\theta$.
At the nodes $z_0$ of $\phi(r)$,
the modulus of the Higgs field vanishes.
Therefore,
these nodes correspond to the locations of monopoles.

The Euler-Lagrange equations possess the discrete symmetry
\begin{equation}
z\rightarrow -z \ , \  H_1\rightarrow -H_1 ,  \ H_2\rightarrow H_2 ,  \   
H_3\rightarrow H_3 , \ \ 1-H_4\rightarrow -(1-H_4) , \ \  
\Phi_1\rightarrow \Phi_1 , \ \Phi_2\rightarrow -\Phi_2 \ . 
\end{equation}
When the solutions possess the same symmetry, we expect
for each node $z_0$ on the positive $z$-axis 
a node $-z_0$ on the negative $z$-axis,
i.~e.~the monopoles come in pairs. 
For a solution with total magnetic charge $Q=0$,
then half of the monopoles carry negative magnetic charge 
and half of them positive magnetic charge.
In the simplest non-trivial case 
the solution then describes a monopole-antimonopole pair.

For such a monopole-antimonopole pair we expect
a magnetic dipole field for the asymptotic gauge potential.
Deriving the asymptotic behaviour of the gauge potential,
we find that
the gauge field function $H_3$ decays like $O(1/r)$ at infinity,
while the other gauge field functions decay exponentionally. 
In particular, in the gauge where the 
Higgs field is asymptotically constant \cite{note1}
i.~e.~$\Phi \rightarrow \tau_3$,
\begin{equation} 
H_3 = \frac{C_{\rm \bf m}}{r} \sin\theta 
\ , \end{equation}
which leads to the asymptotic gauge potential 
\begin{equation}
A_i = C_{\rm \bf m}\frac{(\vec{e}_z \times \vec{r})_i}{r^3} \tau_3 \ ,
\end{equation}
representing indeed a magnetic dipole field.

\boldmath
\section{Magnetic charges of the $Q=0$ configuration}
\unboldmath

As a consequence of the vanishing topological number, $Q=0$,
the configuration we are considering
carries zero net magnetic charge, $m=0$.
In the following we demonstrate, that this field configuration
still possesses magnetic charges.

Let us parameterize the Higgs field as
\begin{equation}
\Phi = \tilde{\Phi}_1 \tau_\rho + \tilde{\Phi}_2 \tau_3 \ ,
\end{equation}
with
\begin{equation}
\tilde{\Phi}_1 = \sin 2\theta \Phi_1 + \cos 2\theta \Phi_2 \ , \ \ \
\tilde{\Phi}_2 = \cos 2\theta \Phi_1 - \sin 2\theta \Phi_2 \ , 
\end{equation}
and define the normalized Higgs field by 
\begin{equation}
\hat{\Phi} = \cos\alpha \tau_\rho + \sin\alpha \tau_3 \ , 
\end{equation}
with
\begin{equation}
\tilde{\Phi}_1 = |\Phi| \cos\alpha \ , \ \ \
\tilde{\Phi}_2 = |\Phi| \sin\alpha \ , \ \ \
|\Phi|  = 
\sqrt{\tilde{\Phi}_1^2+\tilde{\Phi}_2^2} =\sqrt{\Phi_1^2+\Phi_2^2} \ . 
\end{equation}
Thus, $\hat{\Phi}$ maps any closed surface ${\cal S}$ 
in coordinate space to a 2D sphere in isospin space.
We define the degree of the map as
\begin{equation}
\rho^{({\cal S})} = \frac{-i}{2V({\cal S})}\int_{{\cal S}}
             {Tr\left\{\hat{\Phi} d\hat{\Phi}\wedge d\hat{\Phi}\right\}} \ ,
\end{equation} 
where $V({\cal S})$ is the volume of the surface ${\cal S}$.

Let us first calculate the degree of the map of a 2D sphere $S^2$ 
of radius $r$ centered at the origin. We find 
\begin{equation} 
\rho^{(S^2)}(r) = -\frac{1}{4\pi}
   \int_0^{2\pi}\int_0^{\pi}\partial_\theta \sin\alpha d\theta d\varphi
 = - \left. \frac{1}{2}\sin\alpha \right|_{\theta=0}^{\theta=\pi} \ .    
\end{equation}
We now anticipate that the modulus of the Higgs field possesses
two zeros located at $z_0$ and $-z_0$ on the positive and negative 
$z$-axis, respectively.
Then the function $\alpha$ possesses discontinuities on the $z$-axis at 
$z_0$ and $-z_0$, 
i.~e.~$\alpha = -3\pi/2$ for $ -\infty < z < -z_0$, 
$\alpha = -\pi/2$ for $ -z_0 < z < z_0$ 
and  $\alpha = \pi/2$ for $z_0 < z < \infty$.
For any $r$ we find then $\rho^{(S^2)}(r)=0$, 
thus the map $S^2 \rightarrow S^2$ 
can be contracted to the trivial map.

Next we consider the integral over the half sphere 
\begin{equation}
H^2_+ := \left\{ (r,\theta,\varphi)\ \ | \ \ r \  {\rm fixed} \ , 
 \theta \in [0,\frac{\pi}{2}]  \ , \varphi \in [0,2\pi] \right\}\ .
\end{equation}
Note that the boundary of $H^2_+$ is a circle in the $xy$-plane,
\begin{equation}
\partial H^2_+ = \left\{ (r,\theta,\varphi)\ \ | \ \ r \  {\rm fixed} \ , 
 \theta =\frac{\pi}{2}  \ , \varphi \in [0,2\pi] \right\}\ .
\end{equation}
On the boundary the map $\hat{\Phi}$ is 
constant, $\hat{\Phi}(r,\pi/2,\varphi) = -\tau_3$. 
We now compactify $H^2_+$ to a 2D sphere $S^2_+$, by identifying the 
boundary of $H^2_+$ with the south pole of $S^2_+$.
Calculating the degree of the map we find 
\begin{equation} 
\rho^{(S^2_+)}(r) = -\frac{1}{4\pi}
   \int_0^{2\pi}\int_0^{\pi/2}\partial_\theta \sin\alpha d\theta d\varphi
 = -\left. \frac{1}{2} \sin\alpha \right|_{\theta=0}^{\theta=\pi/2}  
= \Theta(r-z_0)\ .    
\end{equation}
Thus, the degree of the map vanishes if the location of the zero $z_0$
is not inside the sphere $S^2_+$ and equals one otherwise. 
An analogous calculation for the lower half sphere $H^2_-$ leads to 
\begin{equation}
\rho^{(S^2_-)}(r) = - \Theta(r-z_0)\ . 
\end{equation}

Let us now compare with the magnetic charges $m$
in the upper and lower half spaces.
To this end, we consider
the `t Hooft electromagnetic field strength tensor 
\begin{equation}
{\cal F}_{\mu\nu} =
Tr\left\{ \hat{\Phi} F_{\mu\nu} 
 -\frac{i}{2} \hat{\Phi} D_\mu \hat{\Phi} D_\nu \hat{\Phi} \right\} \ ,
\end{equation} 
or equivalently
\begin{equation}
{\cal F}_{\mu\nu} =
-\frac{i}{2} Tr\left\{\hat{\Phi} 
\partial_\mu \hat{\Phi} \partial_\nu \hat{\Phi} \right\}
+ Tr\left\{ \partial_\mu(\hat{\Phi}A_\nu)
-\partial_\nu(\hat{\Phi}A_\mu)\right\} \ .
\label{emtH}
\end{equation} 
The magnetic charge inside a 
closed surface ${\cal S}$ can be expressed as
\begin{equation}
m = \frac{1}{V({\cal{S})}} 
\int_{{\cal S}}{ {\cal F}_{\mu\nu} }dx^\mu dx^\nu \ .
\end{equation}
Note, that the integration of the first term in Eq.~(\ref{emtH}) 
leads to the degree of the map.

For the upper half space 
we define the closed surface ${\cal S}_+ = H^2_+\cup D^2 $, 
where $D^2$ is the disk of radius $r$ in the $xy$-plane 
centered at the origin.
Taking into account the correct orientation of the surface element,  
we find
\begin{eqnarray}
m^{({\cal S}_+)} & = & \frac{1}{V({\cal{S}_+)}}\left(
\int_{H^2_+}{{\cal F}_{\theta\varphi} }d\theta d\varphi
   -\int_{D^2}{{\cal F}_{r\varphi} }dr d\varphi \right)
\nonumber\\
 & = &  \rho^{(H^2_+)} 
 + \frac{1}{4\pi}\left(\int_{H^2_+}{
  Tr\left\{ \partial_\theta(\hat{\Phi}A_\varphi)\right\}}  
                 d\theta d\varphi
 -\int_{D^2}{
  Tr\left\{ \partial_{r'}(\hat{\Phi}A_\varphi)\right\}}  
                 dr' d\varphi \right)
\nonumber\\
 & = &  \rho^{(H^2_+)} 
   +\frac{1}{4\pi}\left(\int_0^{2\pi}
\left. 
Tr\left\{\left(\hat{\Phi}A_\varphi\right)
|_r \right\}\right|_{\theta=0}^{\theta=\frac{\pi}{2}} d\varphi          
   -\int_0^{2\pi}
\left. Tr\left\{\left(\hat{\Phi}A_\varphi\right)
 |_{\theta=\frac{\pi}{2}}\right\}
   \right|_{r'=0}^{r'=r}  d\varphi     \right)       
                 \ , 
\end{eqnarray}
where we have used that $Tr\{\hat{\Phi}A_\theta\}=Tr\{\hat{\Phi}A_r\}=0$,
and that the first term in  Eq.~(\ref{emtH}) does not 
contribute to the integration over the disk $D^2$.
 From symmetry considerations we know that $A_\varphi|_{\theta=0}=0$, 
$A_\varphi|_{\theta=\frac{\pi}{2}}= H_3 \tau_3$ and $H_3(r=0)=0$.
Hence we find (see also Appendix A)
\begin{eqnarray}
m^{({\cal S}_+)} & = & \rho^{(H^2_+)} 
+\frac{1}{4\pi}\left(\left.  
2\pi(2 \sin\alpha H_3)\right|_{\theta=\frac{\pi}{2}}
-\left.  2\pi(2 \sin\alpha H_3)\right|_{\theta=\frac{\pi}{2}}\right)
\nonumber\\
 & = & \rho^{(H^2_+)}  
                 \ . 
\end{eqnarray}
Analogously, for the lower half space we find
\begin{equation}
m^{({\cal S}_-)} = \rho^{(H^2_-)} \ .
\end{equation}

These calculations show indeed,
that the configuration possesses two magnetic charges with opposite sign,
located on the positive and negative $z$-axis, respectively. 
In general, for closed surfaces which contain only 
one of the locations of the zeros of the Higgs field, integration 
of the field strength tensor normal to the surface yields a non-vanishing 
magnetic charge.
In constrast, for a surface enclosing both charges,
their contributions compensate,
yielding zero net magnetic charge.

\section{Numerical Results}

The Ansatz Eqs.~(\ref{A_an}), (\ref{Phi_an}) is form invariant under 
abelian gauge transformations 
${\displaystyle {\bf g}= e^{i\Gamma/2 \tau_\varphi}}$,
with the gauge and Higgs field functions transforming as
\begin{eqnarray}
-\frac{H_1}{r} & \rightarrow & -\frac{H_1}{r} + \partial_r \Gamma \ , 
\nonumber \\
2H_2 & \rightarrow & 2H_2 + \partial_\theta \Gamma \ , 
\nonumber \\
\left(H_3 -\frac{\cos2\theta}{2\sin\theta}\right) 
& \rightarrow &
 \cos\Gamma \left(H_3 -\frac{\cos2\theta}{2\sin\theta}\right) 
+\sin\Gamma \left(1-H_4 -\cos\theta\right) \ ,  
\nonumber \\
\left(1-H_4 -\cos\theta\right) & \rightarrow &
 \cos\Gamma \left(1-H_4 -\cos\theta\right) 
-\sin\Gamma \left(H_3 -\frac{\cos2\theta}{2\sin\theta}\right) \ ,
\nonumber \\
\Phi_1 &\rightarrow & \cos\Gamma \Phi_1 + \sin\Gamma \Phi_2   \ , 
\nonumber \\
\Phi_2 &\rightarrow & \cos\Gamma \Phi_2 - \sin\Gamma \Phi_1   \ . 
\end{eqnarray}
To find a unique solution we have to fix the gauge and choose
the condition
\begin{equation}
G_{\rm f} = 
\frac{1}{r^2}\left( r\partial_r H_1 -2 \partial_\theta H_2\right) =0 \ .
\end{equation}
The set of partial differential equations is then obtained 
from the Lagrangian Eq.~(\ref{lag1}) with the gauge fixing term
$\xi G_{\rm f}^2$ added, where $\xi$ is a Lagrange multiplier.

This set of partial differential equations is solved numerically
subject to the following boundary conditions, 
which respect finite energy and finite 
energy density conditions as well as regularity and symmetry requirements. 
These boundary conditions are at the origin
\begin{equation}
H_1(0,\theta)=H_3(0,\theta)=0\ , \ \ \ \ H_2(0,\theta)=H_4(0,\theta)=1 \ , 
\label{BCH0}
\end{equation}
\begin{equation}
\sin 2\theta \Phi_1(0,\theta)+\cos 2\theta \Phi_2(0,\theta) = 0 \ , \ \ \ \ 
\partial_r (\cos 2\theta \Phi_1(0,\theta)
 -\sin 2\theta \Phi_2(0,\theta)) = 0 \ ,
\label{BCP0}
\end{equation}
at infinity
\begin{equation}
H_1(\infty,\theta)=H_2(\infty,\theta)=0\ , \ \ \ \ 
H_3(\infty,\theta)=\sin\theta \ , \ \ 
\left(1-H_4(\infty,\theta)\right)=\cos\theta
\end{equation}
\begin{equation}
\Phi_1(\infty,\theta)=1\ , \ \ \ \ \Phi_2(\infty,\theta)=0 \ ,
\end{equation}
and on the $z$-axis
\begin{equation}
H_1(r,\theta=0,\pi) = H_3(r,\theta=0,\pi) 
 = \partial_\theta H_2(r,\theta=0,\pi)
 = \partial_\theta H_4(r,\theta=0,\pi) = 0 \ ,
\end{equation}
\begin{equation}
\Phi_2(r,\theta=0,\pi)
 = \partial_\theta \Phi_1(r,\theta=0,\pi)=0\ .   
\end{equation}

We have constructed monopole-antimonopole solutions
for a large range of values of the Higgs coupling constant $\lambda$.
The numerical calculations were performed with the software package 
CADSOL/FIDISOL, based on the Newton-Raphson method \cite{schoen}.
For vanishing Higgs coupling constant
the monopole-antimonopole solution 
corresponds to a non-Bogomol'nyi BPS solution,
for which our results are in good agreement 
with those of Ref.~\cite{Rueb}.

In Table 1 we present the normalized energy of the solutions 
$E/4\pi\eta$ for several values of $\lambda$ 
and compare with the energy $E_{\rm inf}/4\pi\eta$
of a monopole-antimonopol pair with infinite separation,
corresponding to twice the energy of a charge-1 monopole. 
For all values of $\lambda$ in Table 1 the energy of the 
monopole-antimonopole solution is less than the energy of a
monopole-antimonopole pair with infinite separation.
%but the difference of the energies $E_{\rm inf}$ and $E$
%decreases monotonically with increasing $\lambda$.
\begin{table}[!h]
\begin{center}
\begin{tabular}{|c|cc|c|c|c|} 
 \hline 
$\lambda$ & $E/4\pi\eta$ &  
 $E_{\inf}/4\pi\eta$ &$d$& $\phi_0$ & $C_{\rm \bf m}$ \\ 
 \hline 
0    & 1.697 & 2.000 & 4.23 & 0.328 & 2.36 \\
0.001& 1.830 & 2.053 & 3.48 & 0.381 & 2.07\\
0.01 & 2.015 & 2.204 & 3.34 & 0.489 & 1.84\\
0.1  & 2.330 & 2.498 & 3.26 & 0.791 & 1.71\\
0.2  & 2.442 & 2.613 & 3.24 & 0.886 & 1.69\\ 
0.5  & 2.596 & 2.776 & 3.11 & 0.961 & 1.62 \\ 
1.0  & 2.713 & 2.900 & 3.0  & 0.986 & 1.57\\ 
10.0 & 3.042 & 3.241 & 3.0  & 0.9996 & 1.55\\ 
 \hline 
\end{tabular}
\end{center}
{\bf Table 1}\\
The energy of the monopole-antimonopole solution
as well as the energy of two infinitely separated
$Q=\pm 1$ monopoles, the distance $d$ between the locations
of the monopole and antimonopole,
the modulus of the Higgs field at the origin, $\phi_0$, 
and the dimensionless dipole moment $C_{\rm \bf m}$ 
are given for several values of the Higgs coupling constant $\lambda$.
\end{table}

In Fig.~1 we exhibit the 
modulus of the Higgs field $|\Phi(\rho,z)|$ as a function
of the coordinates $\rho=\sqrt{x^2+y^2}$ and $z$ for $\lambda=0$ 
and $\lambda=1$.
The zeros of $|\Phi(\rho,z)|$ are located on the positive and negative
$z$-axis at $\pm z_0 \approx 2.1 $ for $\lambda=0$ 
and at $\pm z_0 \approx 1.5 $ for $\lambda=1$.
The distance $d$ of the two zeros of the Higgs field decreases 
monotonically with increasing $\lambda$,
as seen in Table 1.

Asymptotically $|\Phi(\rho,z)|$ approaches the value one.
For $\lambda>0$ the decay of the Higgs field is exponentially.
The value of the modulus of the Higgs field at the origin 
increases monotonically with increasing $\lambda$ (see Table 1). 
While $\phi_0 = 0.328$ for $\lambda=0$,
$\phi_0$ is already close to one for $\lambda =1$.
In the limit $\lambda \rightarrow \infty$ we expect the 
modulus of the Higgs field to be equal to one everywhere,
except for two singular points on the $z$-axis, 
representing the locations of the monopole and antimonopole.
In contrast, the angle $\alpha$ should remain a nontrivial
function in this limit.
This would then be similar to the result found in \cite{KKT} 
for the charge-2 multimonopole.
\clearpage
\begin{figure}[!h]
\centering
\mbox{
\epsfxsize=8.cm\epsffile{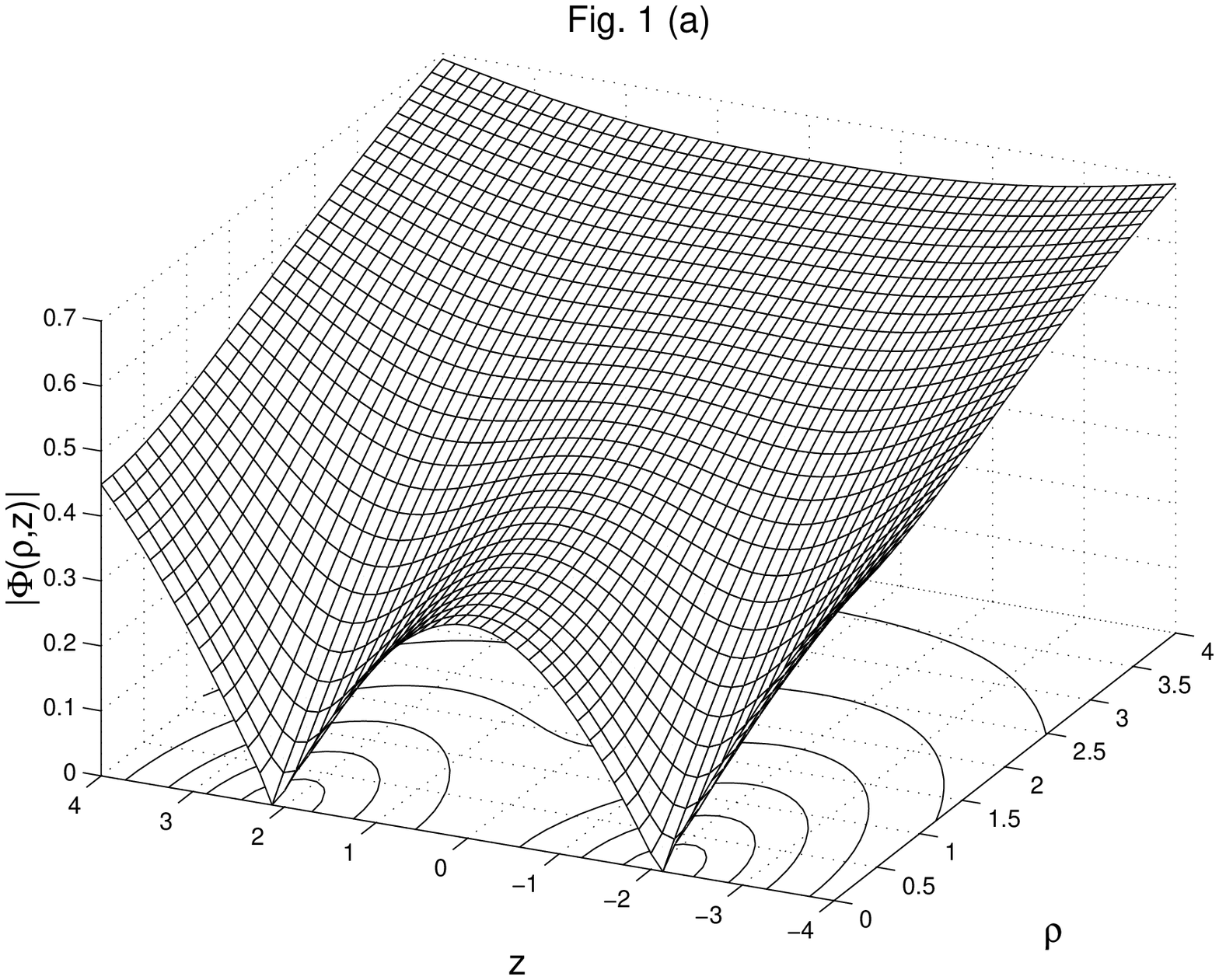}}
\epsfxsize=8.cm\epsffile{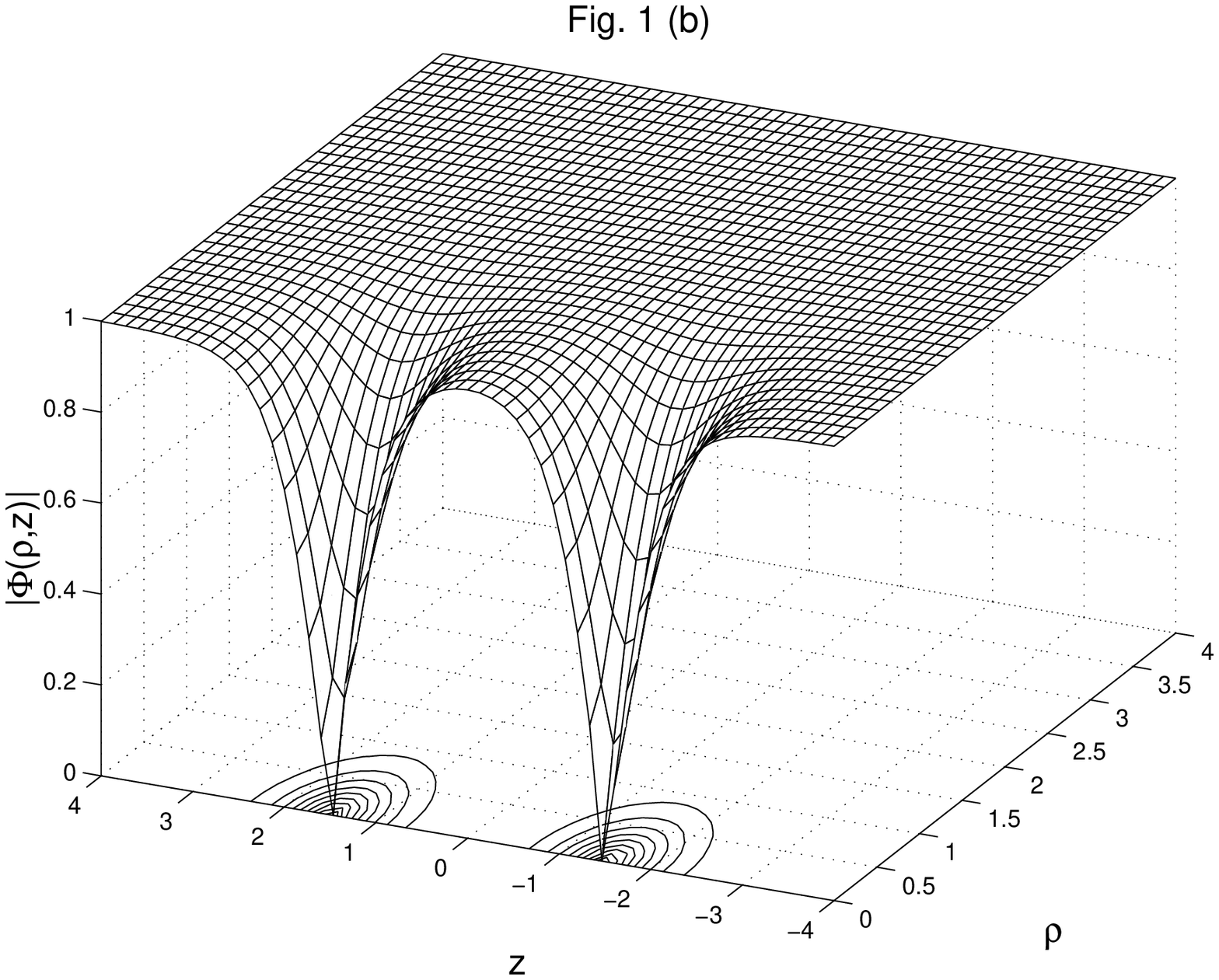}
\caption{
The modulus of the Higgs field as a function of $\rho$ and $z$ for 
$\lambda=0$ (a) and $\lambda=1$ (b) }
\end{figure}

In Fig.~2 we show the energy density of the 
monopole-antimonopole solution as a function
of the coordinates $\rho=\sqrt{x^2+y^2}$ 
and $z$ for $\lambda=0$ and $\lambda=1$.
At the locations of the zeros of the Higgs field
the energy density possesses maxima.
For $\lambda=1$ the maxima are more pronounced compared 
to the case of vanishing $\lambda$.
At large distances from the origin the energy density vanishes 
like $O(r^{-6})$. 
For intermediate distances from the origin the shape of equal energy 
density surfaces looks like a dumb-bell. 
For smaller distances the dumb-bell splits into two surfaces. 

Near the locations of the zeros of the Higgs field 
the equal energy density surfaces 
assume a shape close to a sphere, centered at the 
location of the respective zero. 
This presents further support for the conclusion,
that at the two zeros of the Higgs field 
a monopole and an antimonopole are located,
which can be clearly distinguished from each other, 
and which together form a bound state.
This is in contrast to the axially symmetric 
charge-2 multimonopole solution, where
the individual monopoles cannot be distinguished,
and where, in fact, the Higgs field has only one (double) zero.

\begin{figure}[!h]
\centering
\mbox{
\epsfxsize=8.cm\epsffile{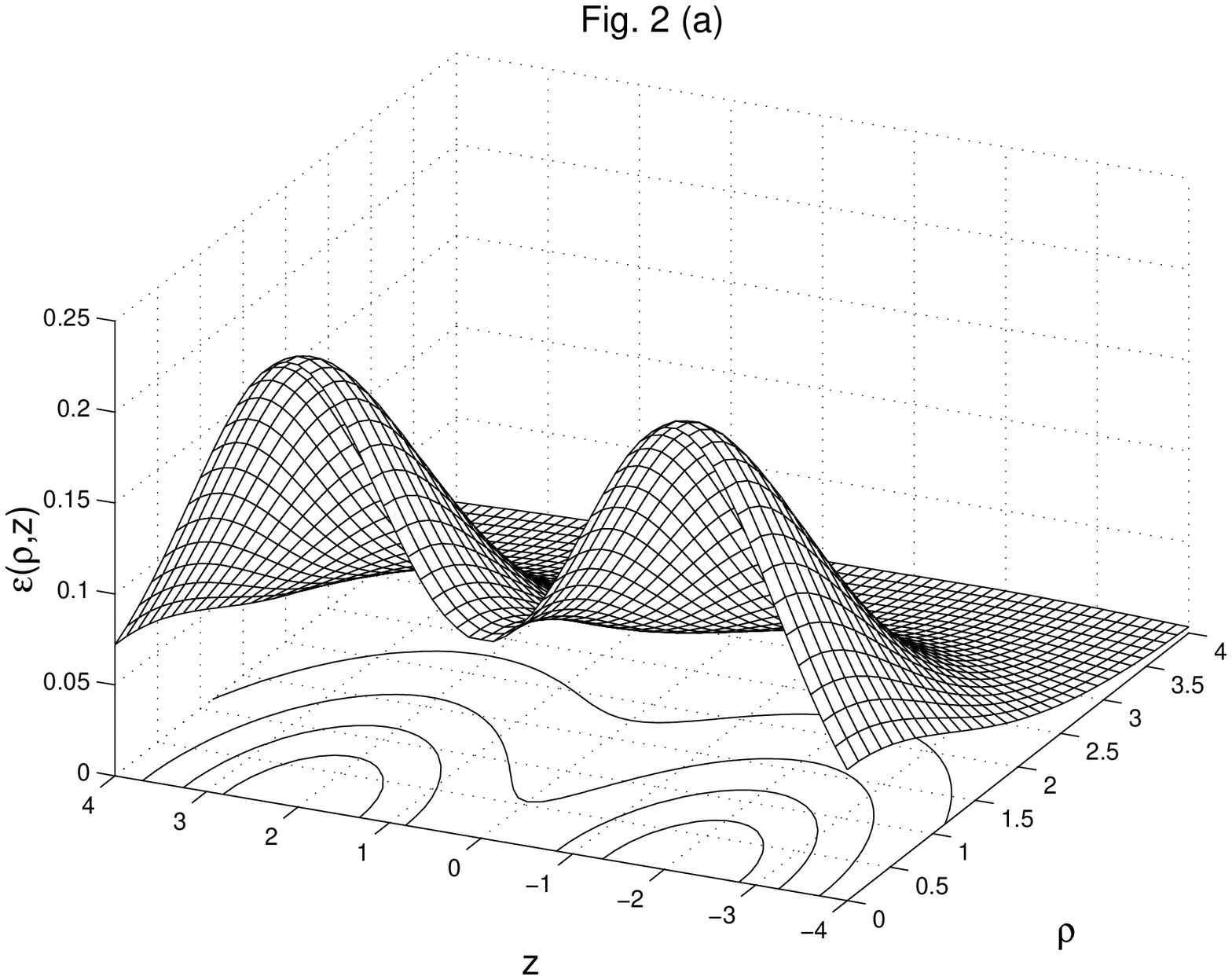}}
\epsfxsize=8.cm\epsffile{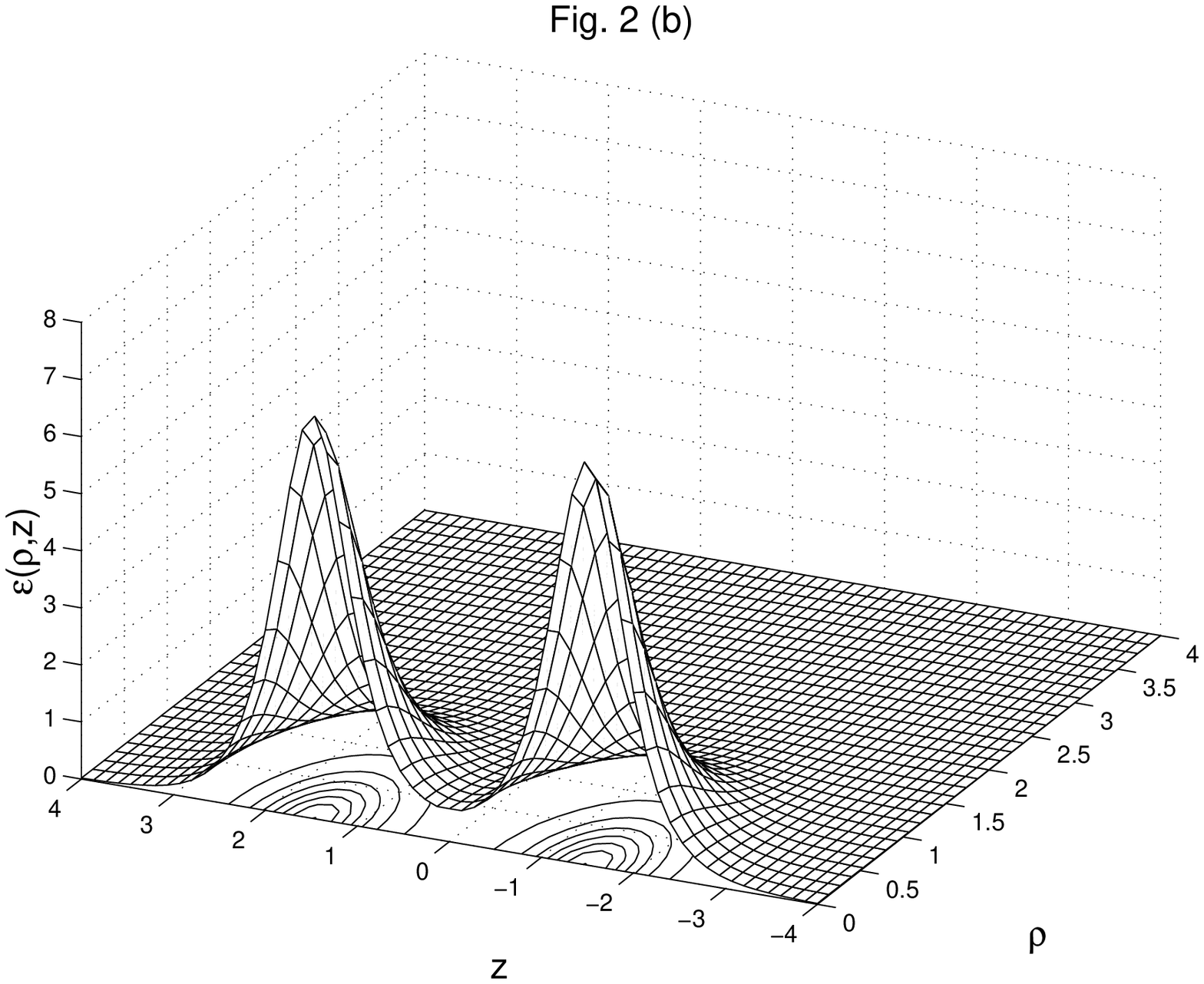}
\caption{
The the dimensionless energy density as a function of $\rho$ and $z$ 
for $\lambda=0$ (a) and $\lambda=1$ (b)}
\end{figure}

Considering finally the electromagnetic properties of the 
monopole-antimonopole solution, we observe, that
the dimensionless dipole moment $C_{\rm \bf m}$ 
decreases monotonically with increasing $\lambda$
(see Table 1).

\section{Conclusions}

We have considered static axially symmetric solutions of the $SU(2)$
Yang-Mills-Higgs model, residing in the vacuum sector.
These solutions represent monopole-antimonopole pairs.
The modulus of the Higgs field possesses zeros
at the locations of the monopole and antimonopole.
Clearly distinguished from each other, 
the monopole and antimonopole together form a bound state,
which carries a magnetic dipole moment and net zero magnetic charge.
However, this bound state is unstable,
corresponding to a saddle point \cite{Taubes,Rueb}.

We have constructed the monopole-antimonopole solutions numerically
for various values of the coupling constant $\lambda$,
representing the strength of the Higgs potential.
With increasing $\lambda$,
%the binding energy $E_{\rm inf}-E$ of the pair,
the energy $E$ of the pair 
and the ratio $E/E_{\inf}$ increase  and
the distance $d$ between monopole and antimonopole as well as
the magnetic dipole moment $C_{\rm m}$ decrease,
while the energy density becomes more localized
around the monopole and antimonopole locations.

In the BPS limit, the $SU(2)$ monopole-antimonopole solution
does not satisfy the first order Bogomol'nyi equations \cite{Rueb}.
Hopefully, the numerical solution will be of help in constructing
this non-Bogomol'nyi solution analytically.
Recently also non-Bogomol'nyi $SU(N)$ BPS solutions, corresponding
to monopole-antimonopole configurations, have been found
\cite{nonB}. These solutions, however, are spherically symmetric.

\vspace{1cm}

{\large \bf Acknowledgments}

B.~K.~was supported by Forbairt grant SC/97-636.

\section{Appendix A}

At first glance it seems to be obvious, that the second term in 
Eq.~(\ref{emtH}) does not contribute to the surface integral, 
because it is the curl of the ``gauge field'' 
$\tilde{A}_\mu = Tr\{\hat{\Phi}A_\mu\}$. However, $\hat{\Phi}$ is not 
continuous on the $z$-axis and introduces a singularity in the 
``gauge field'' $\tilde{A}_\mu$; its curl may contain $\delta$-functions if
the singularity is strong enough. 

To examine the singularity we expand the functions near the singular point
$\vec{r}_s = (0,0, z_0)$,
\begin{eqnarray}
\Phi_1 & = & f_0 +f_2\sin^2\theta \ , \ \ 
\Phi_2 \  = \  f_1\sin\theta \ ,  \ \  \ \ f_0 = c_0( z-z_0 )
\nonumber \\
\tilde{\Phi}_1 & = & \sin\theta (f_1 +2 f_0 ) \ , \ \ 
\tilde{\Phi}_2 =  f_0 +\sin^2\theta (f_2-2(f_0+f_1))
\nonumber \\
H_3 & = & \sin\theta H_{31}  \ , \ \ 
1-H_4 \ = \ H_{40} + \sin^2\theta H_{42} \ ,
\nonumber \\
\end{eqnarray}
where $c_0$, $f_1$, $f_2$, $H_{31}$, $H_{40}$ and $H_{42}$ are constants.
This leads to
\begin{eqnarray}
             & & 
\cos\alpha \ =  \ 
\sin\theta\frac{f_1+2 f_0}{\sqrt{f_0^2 +\sin^2\theta f_1^2}} \ , \ \   
\sin\alpha \ = \ \frac{f_0+\sin^2\theta (f_2-2(f_0+ f_1))}
                   {\sqrt{f_0^2 +\sin^2\theta f_1^2}} \ , \ \   
\nonumber \\
 & & \tilde{A}_x \ = \ \frac{2 y}{z}
  \frac{H_{31}f_0 +H_{40} f_1}
 {\sqrt{z^2f_0^2 +(x^2+y^2)(f_0^2+f_1^2)}}  \ , \ \ 
 \tilde{A}_y \ = \ -\frac{2 x}{z}
  \frac{H_{31}f_0 +H_{40} f_1}{\sqrt{z^2f_0^2 +(x^2+y^2)(f_0^2+f_1^2)}}  \ . 
\nonumber \\
\end{eqnarray} 
Next we define $\bar{z}=z-z_0$ and keep only terms linear 
(quadratic under the square root) in $x$, $y$, $\bar{z}$, 
\begin{equation}
\tilde{A}_x= 2 \frac{y}{z_0}\frac{H_{40} f_1}
 {\sqrt{c_0^2z_0^2 \bar{z}^2 +f_1^2(x^2+y^2)}} 
 \ , \ \  
\tilde{A}_y= -2 \frac{x}{z_0}\frac{H_{40} f_1}
 {\sqrt{c_0^2z_0^2 \bar{z}^2 +f_1^2(x^2+y^2)}} 
 \ .  
\end{equation}  
Now we introduce spherical coordinates 
$x=\bar{r} \sin\bar{\theta} \cos\bar{\varphi}$,
$y=\bar{r} \sin\bar{\theta} \sin\bar{\varphi}$, 
$\bar{z}= \bar{r} \cos\bar{\theta}$ centered at the singular point. 
With respect to these coordinates the components 
of the ``gauge field'' become
\begin{equation}
\tilde{A}_{\bar{r}} = 0 \ , \ \ 
\tilde{A}_{\bar{\theta}} = 0 \ , \ \ 
\tilde{A}_{\bar{\varphi}} = 
-2 \sin^2\bar{\theta}\frac{\bar{r}}{z_0} \frac{H_{40}f_1}
{\sqrt{c_0^2 z_0^2 + \sin^2\bar{\theta}(f_1^2-c_0^2z_0^2)}}
\ . \end{equation}
Then we find for the curl
\begin{equation}
\tilde{F}_{\theta\varphi} = -\frac{2 \bar{r}}{z_0}\partial_\theta
\left\{\sin^2\bar{\theta} \frac{H_{40} f_1}
{\sqrt{c_0^2z_0^2 +\sin^2\bar{\theta}(f_1^2-c_0^2z_0^2)}}\right\} \ .
\end{equation}
Consequently the surface integral 
$\int_0^{2\pi} \int_0^{\pi}\tilde{F}_{\theta\varphi}d\theta d\varphi $
over a sphere centered at the singular point vanishes. 
Thus the singularity of the ``gauge field'' $\tilde{A}_\mu$ is too weak to 
introduce $\delta$-functions in the electromagnetic field strength tensor 
${\cal F}_{\mu\nu}$. 

%%%\newpage
\small{

\end{document}